\documentclass[twocolumn,aps,prb]{revtex4-1}
\usepackage{amsmath,amssymb}
\usepackage{graphicx}
\usepackage{epstopdf}
\usepackage[usenames]{color}

\newcommand{\be}{\begin{equation}}
\newcommand{\ee}{\end{equation}}
\newcommand{\bk}{{{\bf{k}}}}
\newcommand{\bp}{{{\bf{p}}}}
\newcommand{\bK}{{{\bf{K}}}}

\newcommand{\br}{{{\bf{r}}}}

\newcommand{\bea}{\begin{eqnarray}}
\newcommand{\eea}{\end{eqnarray}}
\newcommand{\ra}{\rangle}
\newcommand{\la}{\langle}

\newcommand{\upa}{\uparrow}
\newcommand{\dna}{\downarrow}

\newcommand{\dg}{{\dagger}}
\newcommand{\pdg}{{\phantom\dagger}}

\newcommand{\psib}{\bar{\psi}}
\newcommand{\dk}{d \kern-0.25em \bar~ k}
\newcommand{\dw}{d \kern-0.25em \bar~\omega}
\newcommand{\sgn}{\operatorname{sgn}}

\newcommand{\abs}[1]{\left| #1 \right|} 

\let\goto = \rightarrow 

\begin{document}

\title{Emergent dome of nematic order around a quantum anomalous Hall critical point}
\author{A. M. Cook$^1$, C. Hickey$^1$, A. Paramekanti$^{1,2}$}
\affiliation{$^1$Department of Physics, University of Toronto, Toronto, Ontario, Canada M5S 1A7}
\affiliation{$^2$Canadian Institute for Advanced Research, Toronto, Ontario, M5G 1Z8, Canada}
\begin{abstract}
Motivated by the experimental discovery of the quantum anomalous Hall effect and the interest in
quantum phase transitions of correlated electrons, we consider interaction effects at a quantum anomalous 
Hall critical point. We study a microscopic lattice model of spinful fermions on the triangular lattice which
exhibits a $C=2$ Chern insulator (CI) phase with a quantized anomalous Hall effect, sandwiched
between two normal insulator (NI) phases. The first NI-CI quantum phase transition is driven by 
simultaneous mass inversion of a pair of Dirac fermions, with short range interactions being perturbatively 
irrelevant at the transition. The second CI-NI transition is driven by a quadratic band touching point 
protected by momentum space topology and $C_6$ lattice symmetry. A one-loop renormalization
group analysis shows that short range interactions lead to a single marginally relevant perturbation 
at this transition. We obtain the mean field phase diagram of this model incorporating weak
repulsive Hubbard interactions, finding an emergent dome of nematic order around this CI-NI topological critical 
point. We discuss the crossovers in the Hall conductivity at nonzero temperature, and the Landau theory 
of the quantum and thermal transitions out of the nematic phase. Our results may be relevant to ferromagnetic 
double perovskite films with spin-orbit coupling which have been proposed to host such a Chern transition.
Our work provides perhaps the simplest example of an emergent phase near a quantum critical point.
\end{abstract}
\maketitle

\section{Introduction}
A common theme in strongly correlated electronic matter is the emergence of unexpected orders near quantum phase transitions between
two well-understood phases. This is observed in various heavy fermion materials \cite{Mathur1998,Saxena2000,Park2006,Gegenwart2008} and cuprate 
superconductors \cite{Aeppli1997,vanderMarel2003} where a dome of unconventional superconductivity is believed to be nucleated by critical magnetic
fluctuations around the quantum phase transition associated with the onset of magnetic order. Theoretical work lends strong support to this 
idea, arguing for or explicitly showing that non-Fermi liquid physics and an enhanced tendency to Cooper pairing and superconductivity arise
near certain quantum critical points \cite{Varma1989,Chubukov2003,Berg2012,Metlitski2014}. Similarly,
in the bilayer ruthenate Sr$_3$Ru$_2$O$_7$, nematic order \cite{Borzi2007} is found to appear in a narrow window of 
magnetic field around an underlying metamagnetic quantum critical point \cite{Grigera2001}. 
While there have been various theoretical proposals to understand this nematic \cite{Kee2005,Raghu2009,Conduit2009}, a particularly appealing idea is that it might emerge due to a fermionic variant of order-by-disorder 
physics at the underlying metamagnetic quantum critical point \cite{Green2012}. Can such new and unexpected phases also emerge at topological phase transitions?

This question is partly motivated by recent experiments on (Bi,Sb)$_2$Te$_3$ TI
films doped with magnetic Cr atoms which have reported the first observation of the QAH effect \cite{Xue_Science2013} at temperatures $T \lesssim 0.5$ K.
A tantalizing possibility is that magnetic quantum phase transitions could then drive topological quantum anomalous Hall plateau transitions in such
systems. A more direct motivation
stems from a recent theoretical study \cite{Cook2014} of bilayer films of double perovskite such as Sr$_2$FeMoO$_6$ which
exhibits high temperature ferromagnetism in bulk \cite{Serrate_Review_JPCM2007,Kobayashi1998,Sarma_PRL2000,Erten_SFMO_PRL2011}. For a bilayer grown along the \{111\} direction, we showed that the
combination of spin-orbit coupling and high temperature ferromagnetism in such systems leads to regimes in the phase diagram which support 
a quantum anomalous Hall insulator or, equivalently, a Chern insulator (CI) phase, with Chern number $C=2$. In addition, we found a seemingly 
direct transition from this CI to a topologically trivial normal insulator (NI).
Here, we address the fate of this $C=2$ CI to NI transition taking weak repulsive Hubbard interactions
into account, and discover an emergent dome of nematic order around this underlying quantum critical point as depicted in Fig.~\ref{intphsdiag}.

Our work hinges on the observation that $C=2$ CI to NI transitions in the absence of interactions, which involve a change of Chern number by $2$,  can be 
driven either by simultaneous mass inversion at two Dirac points which are related by inversion symmetry or by a single quadratic band-touching
point (QBTP) protected by lattice symmetries. While the former transition is perturbatively stable against 
interactions, in the latter case interactions in two dimensions (2D) are marginally relevant at the QBTP and can drive unexpected orders around this transition.
Such a QBTP is found in the CI-NI transition discussed in our work on double perovskite bilayers \cite{Cook2014}.

Our study may be viewed as an extension of work on toy models with QBTPs \cite{Sun2009,You2013,Dora2014},
or studies of bilayer graphene (BLG) \cite{Vafek2010, Vafek2014,Zhang2010,Zhang2012,Murray2014}, where the underlying QBTP 
is believed to drive a wide variety of competing phases --- nematic states, layer polarized states, topological insulators, or quantum anomalous Hall 
insulators --- due to marginally relevant electron-electron interactions.
However, in contrast to BLG, our noninteracting model already breaks time-reversal symmetry. Thus, the phases of our model are generically gapped,
and already support a CI phase, with the QBTP only appearing at the CI-NI transition. Furthermore, the absence of multiple valleys and layers
also eliminates the complexity of competing orders inherent to bilayer graphene, \cite{Vafek2010, Vafek2014,Zhang2010,Zhang2012,You2013,Murray2014} leading to a robust phase diagram with an unambiguous nematic phase for weak interactions.
Similar ideas may be of interest to bosonic integer quantum Hall plateau 
transitions  \cite{Grover2013} which naturally involve a change of the Hall conductance plateau by $2 e^2/h$. Our work is also of interest in the broader
context of nematic phases and phase transitions occurring in fermionic quantum Hall systems 
\cite{Balents1996,Joynt1996,Mulligan2010,Abanin2010,Mulligan2011,Sondhi2013,Maciejko2013,You2013} 
and Fermi liquids \cite{Kee2003,Kee2004,MetznerPRB2006,NematicReview2010}. By contrast, QBTPs in three dimensions (3D) are perturbatively 
stable against short range interactions but Coulomb interactions can lead to quantum critical phases \cite{Moon_PRL2013}.

\section{Model}
We consider a tight-binding model of spinful fermions on the triangular lattice with broken time-reversal symmetry, described by the
Hamiltonian 
\be
H =  - \! \sum_{\la ij\ra} t^{\alpha\beta}_{ij}  c^\dg_{i\alpha} c^\pdg_{j\beta} +\Delta \!\! \sum_i (n_{i\upa} - n_{i\dna}) +  
U \sum_i n_{i\upa} n_{i\dna}, \label{eq:Hubbard}
\ee
where $t^{\alpha\beta}_{ij}$
are nearest neighbor hopping amplitudes, $\Delta$ is a local Zeeman splitting, and $U$ is the local Hubbard repulsion.  The spin conserving
hopping matrix elements are chosen to be different; we parametrize them as $t^{\upa\upa}_{ij} = t_1 + t_2$ and $t^{\dna\dna}_{ij} = - t_1 + t_2$.
The spin-flip hopping terms are chosen to be bond dependent, with $t^{\upa\dna}_{i,i+\delta_n} = (t^{\dna\upa}_{i,i+\delta_n})^* = \omega^{n-1} t_3$,
where $\omega={\rm e}^{i 2\pi/3}$, and $\delta_n$ (with $n=1 \ldots 6$) labels the six bonds connecting to nearest neighbors on the triangular 
lattice which make an angle $n \pi/3$ with the $x$-axis.

This model is
motivated by our study \cite{Cook2014} of a \{111\} bilayer of ferromagnetic double perovskites, such as Sr$_2$FeMoO$_6$
or Sr$_2$CrWO$_6$, which have spin-orbit coupled conduction electrons. In bulk double perovskite materials such as Ba$_2$FeReO$_6$,
the spin-orbit coupling impacts the spin dynamics  \cite{Plumb_PRB2013} and leads to the formation of Weyl metals with large anomalous Hall conductivity.
\cite{Cook_PRB2013} The relevant bands for this 
physics are the $j=3/2$ spin-orbit coupled electronic states of the $4d$ or $5d$ transition metal ion (respectively, Mo in Sr$_2$FeMoO$_6$, Re in
Ba$_2$FeReO$_6$, and W in Sr$_2$CrWO$_6$) which get Zeeman split by the ferromagnetic ordering of Fe moments leading to broken
time-reversal symmetry.
In 2D bilayers, where we find evidence of nontrivial Chern bands,
the two spin states in the above Hamiltonian correspond to the two lowest Zeeman-split sublevels of spin-orbit coupled $j=3/2$ atomic states of the 
triangular Mo sublattice. The ferromagnetic ordering of Fe moments which breaks 
time-reversal symmetry also leads to spin-dependent hopping, and complex spin-flip hopping 
amplitudes $t^{\alpha\beta}_{ij}$, dictated by the $j=3/2$ wave functions of Mo. 

In momentum space, defining $\Psi^\dg(\bk) = (c^\dg_\upa(\bk),c^\dg_\dna(\bk))$, the kinetic energy term is
$\sum_\bk \Psi^\dg(\bk) H_0(\bk) \Psi(\bk)$, with
\bea
H_0(\bk)= \begin{pmatrix} A_\bk & D_\bk \\ 
D^*_\bk &  B_\bk \end{pmatrix}. \label{Eq:H0}
\eea
The coefficients in the matrix are
\bea
A_\bk&=&-2 (t_1 + t_2) (\cos k_a + \cos k_b + \cos k_c) + \Delta \\
B_\bk&=& + 2 (t_1 - t_2) (\cos k_a + \cos k_b + \cos k_c) - \Delta \\
D_\bk &=& -2 t_3 (\cos k_a + \omega \cos k_b +\omega^2 \cos k_c).
\eea
Here, we have defined $k_a=k_x$, $k_b=(-k_x/2+k_y \sqrt{3}/2)$, $k_c=(-k_x/2-k_y \sqrt{3}/2)$.

\section{Noninteracting phase diagram}

In the absence of interactions, $U=0$, and for $\abs{ t_2}  < t_1$, the model in Eq.~(\ref{Eq:H0}) exhibits NI and CI 
phases as we vary $\Delta$. As shown in Fig.~1,
for $\Delta < \Delta^0_{c1} = - 3 t_1$ and for $\Delta > \Delta^0_{c2} = 6 t_1$, the ground state is a NI.
For $ \Delta^0_{c1} < \Delta < \Delta^0_{c2}$, it is a $C= \pm 2$ CI. For $\abs{t_2}  > t_1$, it also supports
metallic phases where individual bands have zero Chern number (normal metal, NM) or nonzero Chern number $C=\pm 2$ 
(Chern metal, CM); we will not consider these metallic regimes 
in this paper, focusing instead on the NI-CI and CI-NI transitions for $\abs{ t_2 } < t_1$.

The NI-CI phase transition at $\Delta^0_{c1}$ occurs via a mass gap closing at the Brillouin zone corners $\pm \bK$. Expanding the Hamiltonian matrix around
the critical point with
$\Delta = \Delta^0_{c1}+r$ and momenta $\bk = \pm \bK + \bp$, and dropping constants, we find the low energy Hamiltonian 
around $\pm \bK$ to be
\bea
\! H^{\rm low}_{\pm \bK} \!=\! \begin{pmatrix} - \frac{3}{4} \left( t_2 + t_1 \right) p^2 + r \!&& \mp \frac{3\sqrt{3}}{2} t_3 \left( p_x +  i p_y \right)  \\
								\mp \frac{3\sqrt{3}}{2} t_3 \left( p_x -  i p_y \right)  \!&& - \frac{3}{4} \left( t_2  -  t_1 \right) p^2 - r
\end{pmatrix}.
\eea
The critical point corresponding to $r=0$ thus supports a pair of massless, 
linearly dispersing, Dirac fermions with energy $\pm \frac{3\sqrt{3}}{2} t_3 |\bp|$. The
simultaneous closing of the gap at opposite Dirac points is protected by inversion symmetry.
Upon going from the NI to the CI, the Chern number changes by $2$ due to
each Dirac cone having ``vorticity'' 1, so that mass inversion at the Dirac points leads to a 
momentum space skyrmion with winding number 2 in the CI phase. 

\begin{figure}[t]
\includegraphics[scale=0.5]{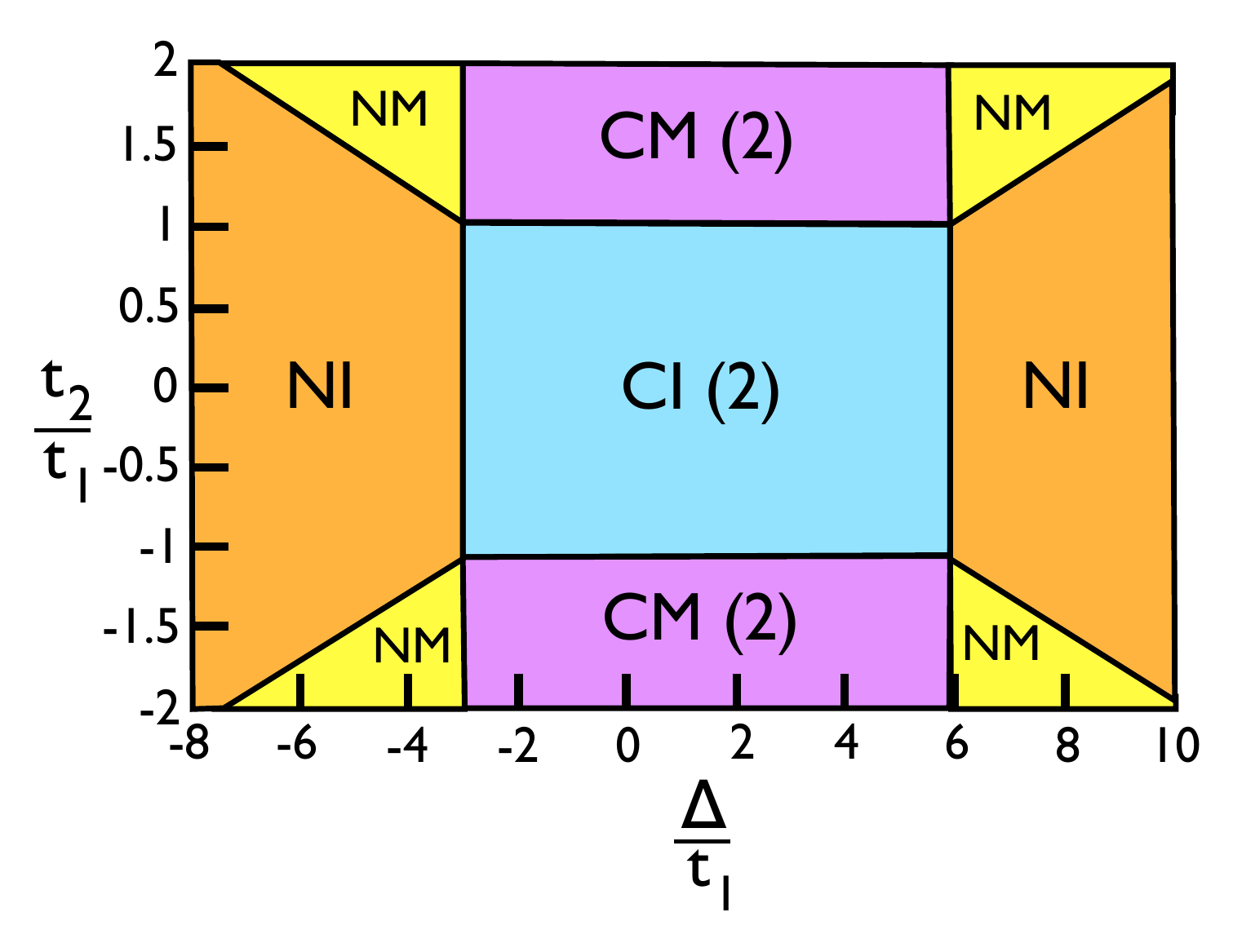}
\caption{\label{nonintphsdiag} Phase diagram for the non-interacting Hamiltonian as a function of $t_2$ and $\Delta$, at a fixed value of $t_3=t_1$.  
Here, phases are CM (Chern metal), NM (normal metal), NI (normal insulator), and CI (Chern insulator).  Non-trivial Chern numbers for the lower band are shown in parentheses.}
\end{figure}

At $\Delta^0_{c2}=6 t_1$, the CI-NI transition occurs via a mass gap 
closing at the $\Gamma$ point.
We can tune near to this critical point and expand the
Hamiltonian for small momenta $\bk$ around the $\Gamma$ point. Dropping constants and setting $\Delta = \Delta^0_{c2}+r$, this leads to 
\be
H^{\rm low}_\Gamma = \begin{pmatrix}  \frac{3}{2} \left( t_2 \!+\! t_1 \right) k^2 + r  &&  \frac{3}{4} t_3 \left(k_x-i k_y \right)^2 \\
\frac{3}{4} t_3 \left( k_x + i k_y \right)^2 &&   \frac{3}{2} \left( t_2 \!-\! t_1 \right) k^2 - r
\end{pmatrix}.
\ee
At $r=0$,
this Hamiltonian supports a quadratic band touching at the $\Gamma$ point. This band touching point
is protected by the $C_6$ lattice symmetry. Upon going from the CI to the NI, the Chern number 
changes by $2$ due to the quadratic band touching point having ``vorticity'' 2, so that mass inversion at the 
$\Gamma$ point unwinds the momentum space skyrmion with winding number 2 in the CI phase
leading back to a normal insulator.

Note that at the CI-NI or NI-CI critical points, where $r\!=\!0$, time-reversal symmetry is only broken ``weakly'', in the sense that there 
are emergent ``Kramers doublets'' at the band touching momenta. However, the dispersion 
away from these momenta has knowledge of the broken time-reversal, leading to coefficients of all three Pauli matrices being nonzero.

\section{RG analysis of interactions}

We next turn to the effect of the short range repulsion, on the above noninteracting 
phase diagram. We can do a renormalization group treatment of interactions, setting
$Z = \int D\psi\psib {\rm e}^{-(S_0 + S_{\rm int})}$ with
\bea
S_0 \!\! &=&\!\!  \int_0^\beta d\tau \sum_{\bk} \psib_{\bk\alpha}(\tau) [\partial_\tau \delta_{\alpha\beta} + H_0(\bk)]_{\alpha\beta} \psi_{\bk\beta}(\tau), \\
S_{\rm int} \!\! &=&\! u \int_0^\beta d\tau \sum_i \psib_{i\upa} (\tau) \psib_{i\dna} (\tau) \psi_{i\dna}(\tau) \psi_{i\upa} (\tau),
\eea
where we have assumed implicit sums on $\alpha,\beta$. The local coupling $u$ in the continuum emerges from the 
Hubbard repulsion $U$ on the lattice.

At $\Delta^0_{c1}$, the critical theory has massless Dirac fermions, and short-range interactions are thus perturbatively 
irrelevant near this transition. Weak repulsive interactions will thus preserve this direct CI-NI transition.

To understand the effect of interactions near $\Delta^0_{c2}$, we
Fourier transform the action (at zero temperature), which leads to
\bea
\!\!\!\! S_0 \!\! &=& \!\! \int_0^\Lambda\!\!\!\!\!\! \dk\!\! \int_{-\infty}^{\infty} \!\!\!\!\!\!\!\! \dw ~ \psib_{\alpha}(\bk,\omega)\left(-i \omega  \delta_{\alpha\beta} \!+\! [H_0(\bk)]_{\alpha\beta} \right) 
\psi_\beta(\bk,\omega) \nonumber \\
S_{\rm int} \! &=& u \int_0^\Lambda \!\!\!\! \{\dk_i\} \int_{-\infty}^{\infty} \!\!\!\!\!\! \{\dw_i\} ~ \psib_\upa(\bk_1,\omega_1) \psib_\dna(\bk_2,\omega_2) \times \nonumber\\ 
&& \psi_\dna(\bk_3,\omega_3) 
\psi_\upa(\bk_1+\bk_2-\bk_3,\omega_1+\omega_2-\omega_3).
\eea
Here $\dk \equiv \frac{d^d\bk}{(2\pi)^d}$, $\dw \equiv \frac{d\omega}{2\pi}$, and the cutoff $\Lambda$ ensures we focus on modes near the $\Gamma$-point.
If we zoom in on the low momentum modes $0 < k < \Lambda {\rm e}^{-\ell}$, and ignore mode-mode interactions, rescaling momenta and frequencies
leads to $du/d\ell = (z-d) u$ where $z$ is the dynamical exponent and $d=2$ is the space dimension.
At $\Delta^0_{c2}$, we have
a quadratic band touching, so $z=2$, and the interactions
are marginal at leading order.

Keeping interactions to 1-loop order near this critical point,
\bea
\frac{d r }{d\ell} &=& 2 r + \frac{u \Lambda^2}{4\pi} \frac{t_1}{\sqrt{t_1^2 + t^2_3/4}}, \\
\frac{d u}{d\ell} &=&  \frac{u^2}{6\pi} \frac{1}{\sqrt{t^2_1 + t^2_3/4}}.
\eea
Since $u>0$, interactions tend to shift the true critical point to $r < 0$, 
so that $\Delta^0_{c2}$ is expected to decrease in the presence of repulsive interactions.
Furthermore, the coupling $u$ is marginally {\it relevant}, similar to that found in earlier work on interaction effects at
QBTPs \cite{Sun2009,Vafek2010, Vafek2014,Zhang2010,Zhang2012,You2013,Murray2014, Dora2014}. We thus expect that while the CI-NI phase boundary bends towards smaller $\Delta$, 
interactions which flow to strong coupling can also nucleate new phases in the vicinity of this putative phase boundary. 
We next turn to a mean field
theory of this new phase.

\section{Emergent nematic order}

Since the only low energy fermion modes near $\Delta^0_{c2}$ are at the $\Gamma$ point, we construct
the simplest mean field theory of the Hubbard interaction $U \sum_i n_{i\upa} n_{i\dna}$,
using a translationally invariant vector order parameter
\be
\vec m = \frac{1}{2 N} \sum_{\bk} \la c^\dg_{\bk\alpha} \vec\sigma_{\alpha\beta} c^{\pdg}_{\bk\beta} \ra.
\ee
Here $N$ is the number of sites, the chemical potential $\mu$ is defined to incorporate Hartree corrections, and 
we have assumed the density stays uniform. Such a mean field treatment is expected to be valid for small $U$.
The Hubbard interaction modifies the
original noninteracting Hamiltonian $H_0$ to the mean field Hamiltonian $H_{\rm mf} (\bk)$, which, dropping constant terms, 
takes the form
\bea
\!\!\!\!\! H_{\rm mf}(\bk)\! =\! \begin{pmatrix} \! A_\bk \!-\! \mu \!-\! U m_z \! & D_\bk \!-\! U (m_x \!-\! i m_y)\\ 
D^*_\bk \!-\! U (m_x\!+\! i m_y) \! &  B_\bk \! -\! \mu \! +\! U m_z \end{pmatrix}.
\eea
Using the eigenvalues and eigenstates of this Hamiltonian, we can recompute $\vec m$, which leads to the following self-consistent equations 
for the particle density $\rho=\la (n_{\br\upa} + n_{\br\dna}) \ra$ and the order parameter $\vec m$, both of which we assume to be spatially uniform. 
Let us define $\xi_\pm(\bk)= \frac{A_\bk + B_\bk}{2} - \mu \pm \Gamma_\bk$,
where
\be
\Gamma_\bk \!=\! \sqrt{\left( \frac{A_\bk\!-\!B_\bk}{2}\!-\! U m_z \right)^2 \!+\! |D_\bk \!-\! U (m_x \! -\! i m_y)|^2}.
\ee
These self-consistency equations, then take the form
\bea
\!\!\!\!\!\! \rho &=& \frac{1}{N} \sum_{\bk,\sigma=\pm} n_F(\xi_\sigma(\bk)) \label{eq:mu} \\
\!\!\!\!\!\! m_z \! &=&\! \frac{1}{2 N} \! \sum_{\bk,\sigma=\pm} \!\! \frac{\frac{A_\bk-B_\bk}{2} - U m_z}{\Gamma_\bk} \sigma n_F(\xi_\sigma(\bk)) \label{eq:mz} \\
\!\!\!\!\!\! m^- \! &=&\!  \frac{1}{2 N} \sum_{\bk,\sigma=\pm} \!\! \frac{D_\bk - U m^- }{\Gamma_\bk} \sigma n_F(\xi_\sigma(\bk)) \label{Eq:m-}
\eea
where $n_F(x)=1/({\rm e}^{x/T}+1)$ is the Fermi function at temperature $T$, and $m^- \!\equiv \! m_x \!-\! i m_y$ in Eq.~(\ref{Eq:m-}).
In these equations, $m_z$ is not a symmetry breaking order parameter; it simply leads to a shift of the underlying $z=2$ quantum critical point 
where fermions become gapless. On the other
hand, $m^- \neq 0$ leads to spontaneous nematic order which breaks lattice rotational symmetry, and simultaneously
splits the QBTP into two massless Dirac points.

\subsection{Nematic phase boundaries}

To obtain an estimate of the regime where nematic order is stabilized by weak interactions, we first obtain the renormalized critical point which is 
obtained by solving Eq.(\ref{eq:mu}) and Eq.(\ref{eq:mz}) self-consistently for $\mu,m_z$. For $\Delta > 0$ we find a self-consistent $m_z < 0$ which renormalizes
the critical point to $(\Delta^0_{c2}+U m_z) < \Delta^0_{c2}$, consistent with predictions from our RG results. For small $U$, this shift is small. 
In this subsection, let us denote $\tilde{r}=\Delta-(\Delta^0_{c2}+U m_z)$ to be the deviation from this renormalized critical point.
Next, we linearize Eq.(\ref{Eq:m-}) for $m^-$ around 
this renormalized critical point to obtain an estimate of the window in the $\Delta$-$T$ phase diagram where the system is unstable to
spontaneous nematic order.

\begin{figure}[t]
\includegraphics[scale=0.30]{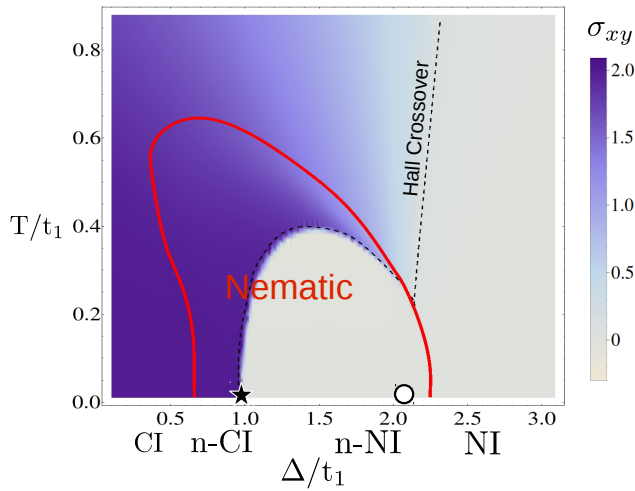}
\caption{\label{intphsdiag}Finite temperature mean field phase diagram showing the nematic dome (marked by solid red line) for the interacting 
Hamiltonian in Eq.~(\ref{eq:Hubbard}) with $t_2/t_1=0.5$, $t_3/t_1=1$, and $U/t_1=8.5$. 
The $T=0$ phases are Chern insulator (CI), nematic Chern insulator (n-CI), nematic normal 
insulator (n-NI), and normal insulator (NI). The Chern transition, i.e. the nCI-nNI transition, is denoted by $\bigstar$. When nematic order is absent,
the underlying quadratic band touching point occurs at $\bigcirc$. The dashed line (``Hall crossover'') marks the closing of the gap in the mean field spectrum; it
signals a crossover in the anomalous Hall conductivity $\sigma_{xy}$ (shown in units of $e^2/h$).}
\end{figure}

\underline{\bf (i) Quantum nematic transition at $T=0$:} 
Taking $m^- \to 0$ in Eq.(\ref{Eq:m-}), with $T=0$, we find
\bea
\frac{1}{U} =  \frac{1}{2} \int \frac{d^2k}{\left(2\pi\right)^2} \frac{  1  } {\gamma_{\bk}} \left(1-\frac{|D_\bk|^2}{2 \gamma_{\bk}^2}\right)   ,
\eea
where 
\be
\gamma_\bk = \sqrt{\left(\frac{A_\bk\!-\!B_\bk}{2}\!-\! U m_z\right)^2 \!+\! |D_\bk|^2},
\ee
and we have set $n_F(\xi_+(\bk)) = 0$, $n_F(\xi_-(\bk)) = 1$. Using the Hamiltonian expanded around the $\Gamma$ point, we obtain
\bea
\abs{ \tilde{r}_c} \approx  \frac { 15 \Lambda^2 t_1  } {   2\left( \sqrt{5} + 2\sgn (\tilde{r})\right)   }  
{\rm e}^{-\frac{ 2\sqrt{5}}{3} \left(  \frac{30 t_1}{U} - \frac{\sgn(\tilde{r})}{\pi}  \right) } ,
\eea
where, we have assumed $\Lambda \gg \abs{\tilde{r}_c}$, and, for simplicity, set $t_3 = t_1$. This leads to an
asymmetric window around the underlying QBTP where nematic order is stabilized, with the onset of nematicity occurring over a
wider range of $\Delta$ for $\tilde{r} < 0$, as we also find from our numerical solution to the mean field equations plotted in Fig.~\ref{intphsdiag}.

\underline{\bf (ii) Chern transition at $T=0$:}
Using the Hamiltonian expanded about the $\Gamma$ point, and writing $\bk$ in polar coordinates as $k e^{i \theta}$, we can estimate the $\bk$ value at which the gap closing occurs for finite $m^-$
\bea 
k^2 = \frac{4 U}{3 t_3} \abs{m^-} \,\, , \, \, \theta = \eta/2 \,,
\eea
where $m^- = \abs{m^-} e^{i \eta}$. For the gap to close we thus require 
\bea
\tilde{r} = - \frac{2 U}{t_1 t_3} \abs{ m^- } .
\eea
For $t_1$, $t_3 >0$ this ensures that $\tilde{r} <0$ and hence the gap closing occurs to the left of the underlying QBTP, as shown in Fig. \ref{intphsdiag}.
A stronger nematic order thus opens up a larger window between the Chern transition and the underlying QBTP.

\underline{\bf (iii) Thermal nematic transition at $\tilde{r}=0$:}
To obtain the mean field estimate for $T_c$, we set $\tilde{r}=0$, and the linearized self-consistent equation for $m^-$ simplifies to
\bea
\!\!\!\! \frac{1}{U} \!=\!  \frac{1}{\alpha}\! \int_0^{\Lambda} \!\! \frac{dk}{2\pi}  \frac{1}{k} \left[  \tanh\left(\frac{ \xi_+}{2T} k^2 \right) \!-\!   \tanh\left(\frac{ \xi_-}{2T} k^2 \right) \right],
\eea
where we have set $\xi_\pm ( \bk ) \equiv \xi_\pm k^2$ around the $\Gamma$ point. The condition $\abs{t_2} < t_1$ ensures that $\xi_- < 0$. Carrying
out the integral, we find
\bea
T_c \sim \sqrt{\abs{\xi_+} \abs{\xi_-}} \Lambda^2 e^{-\frac{2\pi \alpha}{U}},
\eea
where $\alpha = 5\sqrt{5} t_1/3$ for $t_3=t_1$.

\underline{\bf (iv) Nematic dome:}
As shown in Fig.~\ref{intphsdiag}, 
solving the mean field equations at fixed generic values of $(U/t_1,t_2/t_1,t_3/t_1)$ leads to a nematic dome in the $(\Delta,T)$ phase diagram, which is nucleated
around the underlying QBTP. While the mean field theory is expected to be valid at small $U$, we have chosen a value of $U/t_1=8.5$ in order to obtain a 
sizeable nematic
window, which simplifies the numerical computation of the phase diagram. Based on the Landau theory discussed below, which leads to cubic invariants in
the order parameter, 
we expect  mean field transitions out of the nematic to be first order transitions. However, our numerical study finds that the transitions are very weakly
first order in nature, so that our above analytical estimates of the phase boundaries are still useful. The outward splaying of the (red) nematic phase boundary
to the left is due to the extra thermal quasiparticle entropy of the nematic phase, which has a small gap since the Chern transition (at which the single-particle
gap closes) occurs within the nematic phase.

\subsection{Landau theory of the nematic transition}
To gain further insight into the nature of the nematic transition, we construct the Landau free energy for the nematic order parameter. Since we are considering a 
translationally invariant order parameter, we only need to discuss its transformations under the triangular lattice point group symmetries, namely inversion, $C_6$ 
rotations, and reflections.

Inversion leaves $H_0(\bk)$ invariant, thus $\Psi(\bk)$ and hence $m^-$ are left invariant. Under a clockwise rotation by $n \pi/3$
about a site, $D_{\bk} \goto \omega^n D_{\bk}$, hence $m^- \to \omega^n m^-$, and $\Psi^\dg(\bk) = (c^\dg_\upa(\bk), c^\dg_\dna(\bk)) \goto ( c^\dg_\upa(\bk), \omega^n c^\dg_\dna(\bk))$ . This is also seen to leave the gap equation Eq.~(\ref{Eq:m-}) unchanged. 
Under reflection about the $\hat{x}$ axis, $k_a$ is left unchanged while $k_b \leftrightarrow k_c$, which leads to $\Psi (\bk) \goto \Psi^\dg(\bk)$ and $D_\bk \to D^*_\bk$, so that $m^- \leftrightarrow 
m^+$.
Ignoring gradient terms, this leads to the following symmetry allowed terms in the Landau theory of the complex nematic order parameter which we denote by $\phi \sim m^-$
\begin{equation}
f = a |\phi|^2 + b \left(\phi^3 + \phi^{*3} \right) + u |\phi|^4 + \dots ,
\end{equation}
with $u>0$. When $a < 0$, we expect a state with nonzero $|\phi|$, and the cubic term can then be re-written as $2 b |\phi|^3 \cos 3 \theta$, where we have
set $\phi = |\phi| {\rm e}^{i\theta}$. If $b >0$ this favors $\theta=(2 n + 1) \pi/3$ with $n = 0,1,2$ while $b <0$ favors $\theta=2 n \pi/3$ with $n=0,1,2$.
From our microscopic calculation of the energy of the lower band $\xi_- (\bk)$ integrated over the BZ, we find energy minima at $\theta=2n \pi/3$ with $n = 0,1,2$
suggesting that $b <0$. This leads to a splitting of the
QBTP in the dispersion into two Dirac points symmetrically displaced about the $\Gamma$-point along high symmetry lines connecting the 
$\Gamma$ point to opposite BZ corners, breaking $C_6$ rotational symmetry while preserving inversion symmetry.
Going beyond mean field theory, the thermal nematic transition is expected to be in the universality class of the $q=3$ Potts model (equivalently,
clock model) in 2D, which is a continuous phase transition.\cite{WuPottsModel1982} The quantum nematic transition is expected to have a dynamical critical exponent
$z=2$ due to broken time-reversal symmetry which permits linear time-derivative terms, as pointed out and carefully discussed in recent work \cite{You2013} on the field
theory of such a nematic transition. In that case, since $d+z=4$, we might expect mean field theory to be a reasonable guide, and the cubic invariants
in the Landau theory are then likely to drive a first order quantum nematic transition.

\section{Thermal crossover in the Hall conductivity}

At nonzero temperature, 
the NI and CI phases are not well defined since both exhibit a non-quantized anomalous Hall effect. So {\it the only observed sharp order for $T \neq 0$ 
corresponds to the emergent nematic which undergoes a phase transition into the isotropic phase}. Nevertheless, there is a
well-defined 
line in the phase diagram where a band touching occurs in the mean field band structure - there is no singular change in 
thermodynamics or transport across this line, instead a Kubo formula calculation of the Hall conductivity $\sigma_{xy}$ 
shows that it signals a `Hall crossover' from a large to small anomalous Hall effect as depicted in
Fig.~\ref{intphsdiag}. The Hall crossover line at high temperature extrapolates down to $\Delta/t_1 \sim 2$,
which is the location of 
the underlying QBTP in the absence of nematic order. With decreasing temperature, the $\sigma_{xy}$  crossover gets sharper.
The onset of nematic order leads to a significant bending of the Hall crossover line, and it eventually terminates as $T \to 0$, at the true 
CI-NI quantum critical point which is at a smaller value of $\Delta/t_1 \sim 1$, also under the nematic dome. 

\section{Discussion}
We have studied a simple model of a phase transition between a fermionic Chern insulator with $C=2$ and a 
normal insulator with $C=0$. Although the noninteracting theory allows for two types of direct transitions, driven by a pair of simultaneous Dirac band touchings or 
a single quadratic band touching, the latter is shown to become unstable to nematic order in the presence of interactions. This leads to a nematic ``dome'' around this 
topological CI-NI Chern transition.
Interestingly, our weak coupling phase diagram has a true quantum critical point corresponding to the Chern transition as well as ``nearby''
critical point corresponding to the QBTP when nematic order is absent, both occurring under the nematic dome. Furthermore, the Hall crossover line is significantly
affected by the appearance of nematic order. These aspects of the phase diagram are reminiscent of issues discussed in the context of the high
temperature superconductors \cite{Sachdev2010}, namely, multiple types of ``quantum criticality''
thought to occur at different dopings under the superconducting dome in the high temperature cuprate superconductors, and one of which appears to
be located at the point where the pseudogap crossover temperature extrapolates to zero.
In future work, we will discuss the finite temperature crossovers in our model and the strong coupling limit which may yield unusual spin density wave orders,
spin liquids, or other emergent orders also around the Dirac transition.
In conclusion, our model yields perhaps the simplest example of an emergent order near a 2D quantum 
critical point, and may thus lend useful insights into such emergent phases of correlated electrons.

\acknowledgments
We thank E. Fradkin and Y. You for extremely useful comments and illuminating discussions.
This research was funded by NSERC of Canada.

\bibliographystyle{apsrev4-1}

%

\end{document}